\documentclass[a4paper,11pt]{article}
\pdfoutput=1 
\usepackage{jheppub} 
\usepackage[T1]{fontenc} 
\RequirePackage[displaymath]{lineno} 

\usepackage{epsfig}
\usepackage{graphicx}
\usepackage{dcolumn}
\usepackage{bm}
\usepackage{ltablex,booktabs}
\usepackage{overpic}
\usepackage{subfigure}
\usepackage{float}
\usepackage{color}
\usepackage{amsmath}
\usepackage{mathcomp}
\usepackage{mathrsfs}
\usepackage{multirow}
\usepackage{rotating}
\usepackage{amssymb}
\usepackage{gensymb}
\usepackage{amsmath}
\usepackage{tabularx}
\usepackage{epstopdf}

\title{Search for the radiative decay $D^+_s \to \gamma \rho(770)^+$}
\collaborationImg{\includegraphics[width=0.15\textwidth, angle=90]{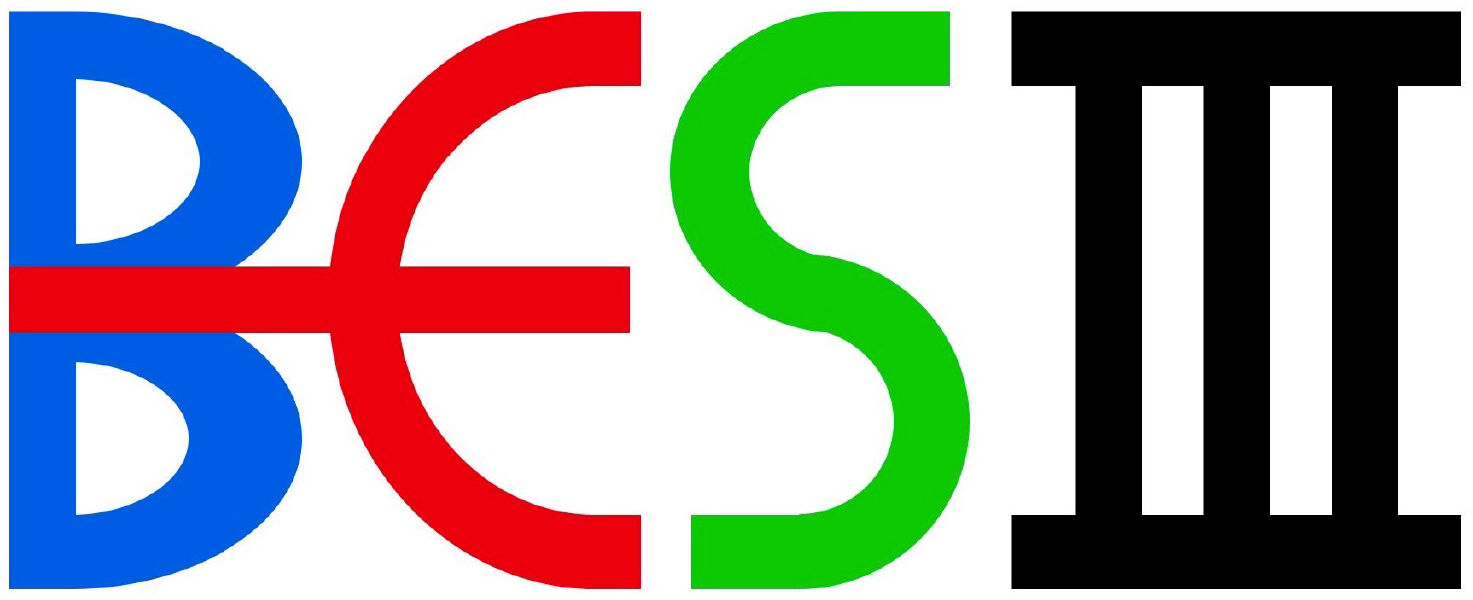}}
\collaboration{The BESIII Collaboration}

\date{\today}
\abstract{Using 7.33 fb$^{-1}$ of $e^+e^-$ collision data samples collected with the BESIII detector at center-of-mass energies between 4.128 and 4.226 GeV, 
we search for the radiative decay $D_s^+\to\gamma\rho(770)^+$ for the first time. 
A hint of $D_s^+\to\gamma\rho(770)^+$ is observed with a statistical significance of 2.5$\sigma$.
The branching fraction of $D_s^+\to\gamma\rho(770)^+$ is measured to be $(2.2\pm0.9_{\rm stat.}\pm0.2_{\rm syst.})\times10^{-4}$, corresponding to an upper limit of $6.1\times$10$^{-4}$ at the $90\%$ confidence level.}
	
\arxivnumber{}
\begin{document}
\maketitle
\flushbottom


\section{Introduction}

In the rare radiative decays of charmed mesons, it has been established that the short-distance interactions of the $c\to u \gamma$ process make a negligible contribution~\cite{Burdman:1995te, CLEO:1998mtp}. Consequently, long-distance, non-perturbative processes dominate these decays, potentially enhancing branching fractions~(BFs) up to $10^{-3}$~\cite{Bajc:1995ys, Fajfer:1997bh, Altmannshofer:2022hfs, Fajfer:1998dv, Belle:2003vsx}.
Therefore, measurements of BFs of these decays can be used to test the quantum chromodynamics~(QCD)-based calculations of long-distance dynamics~\cite{Belle:2016mtj, BaBar:2008kjd}. 
Theoretical physicists have conducted analyses on the $C\kern-0.2em P$ asymmetry of $D_s^+\to\gamma\rho(770)^+$ considering the matrix element contributions of the ${\cal O}_8$ operator in the rare decay Hamiltonian~\cite{Lyon:2012fk}. 
The BF of $D_s^+\to\gamma\rho(770)^+$ can be used to examine the predictions regarding $C\kern-0.2em P$ asymmetry in charmed meson decays.

Recent experimental results on rare radiative charm decays offer a chance to improve the theoretical understanding of physics involved in the $c\to u \gamma$ transition~\cite{Fajfer:2015zea}. However, to date, there have been no experimental results for rare radiative decays of $D_s^+$ mesons, such as the Cabibbo-favored decay $D_s^+\to\gamma\rho(770)^+$. 
The BF of this mode is expected to lie within the range of ${\cal O}(10^{-5})-{\cal O}(10^{-3})$, according to the predictions of different models like 
QCD sum rules~\cite{Khodjamirian:1995uc}, light-cone sum rules~\cite{Lyon:2012fk}, the vector-meson dominance~\cite{Burdman:1995te}, the hybrid framework~\cite{Fajfer:1998dv, Fajfer:1997bh} and weak annihilation~\cite{deBoer:2017que}.
The predictions of the various theoretical calculations are quite divergent for the decay mode $D^+_s\to \gamma\rho(770)^+$, which underlines the urgent need for experimental results to discriminate 
among different models.

In this paper, we search for the radiative decay $D_s^+\to\gamma\rho(770)^+$ for the first time, using 7.33~fb$^{-1}$ data collected with the BESIII detector in $e^+e^-$ collision center of-mass energies ($E_{\rm cm}$) between 4.128 and 4.226 GeV, where $E_{\rm cm}$ is the energy of the initial state calculated from the beam energy in the $e^+e^-$ center-of-mass frame.

\section{Detector and data sets}
\label{sec:detector_dataset}
The BESIII detector records symmetric $e^+e^-$ collisions provided by the BEPCII storage ring~\cite{Yu:2016cof}, in the center-of-mass energy range from 2.0 to 4.95 GeV, with a peak luminosity of $1.1\times 10^{33}$ cm$^{-2}$s$^{-1}$ achieved at the center-of-mass energy of 3.773 GeV.
The cylindrical core of the BESIII detector covers 93\% of the full solid angle and consists of a helium-based multilayer drift chamber~(MDC), 
a plastic scintillator time-of-flight system~(TOF), and a CsI(Tl) electromagnetic calorimeter~(EMC), 
which are all enclosed in a superconducting solenoidal magnet providing a 1.0~T magnetic field. 
The solenoid is supported by an octagonal flux-return yoke with resistive plate counter muon identification modules interleaved with steel.
The charged-particle momentum resolution at 1~GeV/$c$ is $0.5\%$, and the ${\rm d}E/{\rm d}x$ resolution is $6\%$ for electrons from Bhabha scattering. 
The EMC measures photon energies with a resolution of $2.5\%$ ($5\%$) at 1~GeV in the barrel (end-cap) region.
The time resolution in the TOF barrel region is 68~ps, while that in the end-cap region is 110~ps. 
The end-cap TOF system was upgraded in 2015 using multi-gap resistive plate chamber technology, providing a time resolution of 60~ps~\cite{Li:2017jpg, Guo:2017sjt, Cao:2020ibk}. 
About $83\%$ of the data in this analysis benefits from the upgrade.

The data samples are organized into four groups, $E_{\rm cm} =$ 4.128 and 4.157 GeV~(I), 4.178 GeV~(II),  four energies from 4.189 to 4.219 GeV~(III), and 4.226 GeV~(IV), that were acquired during the same year under consistent running conditions. 
The integrated luminosities at each energy is given in table~\ref{energe}.
The signal events discussed in this paper are selected from the process $e^+e^-\to D^{*+}_sD_s^{-}$.

\begin{table}[thbp]
\centering
\begin{tabular}{c c c}
\hline
$E_{\rm cm}$ (GeV)~\cite{BESIII:2020eyu, BESIII:2015zbz} &$\mathcal{L}_{\rm int}$ (pb$^{-1}$)~\cite{BESIII:2022dxl} & $M_{\rm rec}$ (GeV/$c^2$)\\
\hline 
4.128 &401.5   &[2.060, 2.150] \\
	 4.157 &408.7   &[2.054, 2.170] \\
	 4.178 &$3189.0\pm0.2\pm31.9$  &[2.050, 2.180] \\
	 4.189 &$570.0\pm0.1\pm2.2$ &[2.048, 2.190] \\
	 4.199 &$526.0\pm0.1\pm2.1$ &[2.046, 2.200] \\
	 4.209 &$572.1\pm0.1\pm1.8$ &[2.044, 2.210] \\
	 4.219 &$569.2\pm0.1\pm1.8$ &[2.042, 2.220] \\
	 4.226 &$1100.9\pm0.1\pm7.0$ &[2.040, 2.220] \\
\hline
\end{tabular}
\caption{\label{energe} The integrated luminosities ($\mathcal{L}_{\rm int}$) and the requirements on the $D^-_s$ recoil mass ($M_{\rm rec}$) for various center-of-mass energies.
	 The definition of $M_{\rm rec}$ is given in eq.~(\ref{eq:mrec}).
   The first and second uncertainties for $\mathcal L$ are statistical and systematic, respectively. The integrated luminosities for the two data samples at $E_{\rm cm} = 4.128$ GeV and $E_{\rm cm} = 4.157$ GeV are estimated by using online monitoring information.}
\end{table}

Inclusive Monte Carlo~(MC) samples, which are 40 times larger than the data sets, are produced in the energy range $E_{\rm cm} = 4.128$ to $4.226$ GeV with a {\sc geant4}-based~\cite{GEANT4:2002zbu} software package, which includes the geometric description of the BESIII detector and the detector response. 
These samples are used to determine the detection efficiencies and estimate backgrounds.
The samples include the production of open charm processes, the initial-state radiation production of vector charmonium(-like) states and the continuum processes incorporated in {\sc kkmc}~\cite{Jadach:2000ir, Jadach:1999vf}. 
All particle decays are modeled with {\sc evtgen}~\cite{Lange:2001uf, Ping:2008zz} using BFs either reported by the PDG~\cite{PDG}, 
when available, or otherwise estimated with {\sc lundcharm}~\cite{Chen:2000tv, Yang:2014vra}. 
Final state radiation from charged particles is incorporated using {\sc photos}~\cite{Richter-Was:1992hxq}. 
For the signal Monte Carlo samples, we generate 3.2 million events for the decay $D_s^+\to \gamma\rho(770)^+$, $\rho(770)^+\to\pi^0\pi^+$, and $\pi^0\to\gamma\gamma$. The $D^+_s\to\gamma\rho(770)^+$ decay is parameterized by helicity amplitudes~\cite{Belle:2003vsx, Belle:2016mtj, BaBar:2008kjd}, resulting in 
an angular distribution of $1-\rm cos^2\theta_{\rm H}$, in which $\theta_{\rm H}$ is the helicity angle between the momentum vector of the particle $\pi^+$ in the $\rho(770)^+$ rest frame and the direction of the $\rho(770)^+$ system in the $D^+_s$ rest frame. 
The decay $\rho(770)^+\to\pi^0\pi^+$ is modeled with VSS~\cite{Lange:2001uf, Ping:2008zz}, which describes the decay
of a vector meson to scalar-scalar mesons. 
The decay $\pi^0\to\gamma\gamma$ is generated 
with a uniform phase-space model~\cite{Jadach:2000ir, Jadach:1999vf}.

\section{Analysis strategy}
The $D_s^{*\pm}D_s^{\mp}\to\gamma(\pi^0)D_s^{\pm}D_s^{\mp}$ pairs are produced by $e^+e^-$ collisions  in the energy range of 4.128 to 4.226 GeV.
This property  enables the study of $D^+_s$ decays utilizing the double-tag (DT) method, which was pioneered by the MARK-III collaboration~\cite{MARK-III:1985hbd}. 
At first, single-tag (ST) $D_s^-$ candidates are selected by reconstructing a $D^-_s$ in five hadronic decay modes: $D_{s}^{-}\to K_{S}^{0}K^{-}$, $D_{s}^{-}\to K^{+}K^{-}\pi^{-}$, $D^{-}_{s}\to K^{+}K^{-}\pi^{-}\pi^{0}$, $D_{s}^{-}\to K_S^0K^{+}\pi^-\pi^-$ and $D^-_s\to K^-\pi^-\pi^+$.
Events in which a signal candidate is reconstructed in the presence of an ST $D_s^-$ meson are denoted as DT events.

The BF of the signal decay is determined by
\begin{eqnarray}\begin{aligned}
	\mathcal{B}(D_{s}^{+} \to \gamma\rho(770)^+) =
	\frac{N^{\rm DT}_{\rm total}}{{B}(\pi^0\to\gamma\gamma)\sum_{\alpha,i}{N^{\rm ST}_{\alpha,i}}\epsilon^{\rm DT}_{\alpha,i}/\epsilon^{\rm ST}_{\alpha,i}},
	\label{abs:bf}
\end{aligned}\end{eqnarray}
where $\mathcal{B}(\pi^0\to\gamma\gamma)$ is the BF of $\pi^0\to\gamma\gamma$ in the $\rho^+\to\pi^+\pi^0$ decay. $N_{\rm total}^{\rm DT}$ is the summed number of the DT yields for all four sample groups. 
$\epsilon^{\text{ST}}_{\alpha,i}$ is the ST efficiency to reconstruct the ST mode $i$ in the sample group $\alpha$, and $\epsilon_{\alpha,i}^{\text{DT}}$ is the DT efficiency for reconstructing the ST mode $i$ and the signal decay mode in the sample group $\alpha$.
Charge-conjugated modes are implicitly considered throughout this paper.

\section{Event selection}
\label{ST-selection}
Charged tracks detected in the MDC are required to be within a polar angle ($\theta$) range of $|\rm{cos\theta}|<0.93$, where $\theta$ is defined with respect to the $z$-axis, which is the symmetry axis of the MDC. 
For charged tracks not originating from $K_S^0$ decays, the distance of closest approach to the interaction point~(IP) is required to be less than 10~cm along the $z$-axis, $|V_{z}|$, and less than 1~cm in the transverse plane, $|V_{xy}|$.
Particle identification~(PID) for charged tracks combines the measurements of d$E$/d$x$ in the MDC and the time of flight in the TOF to form likelihoods $\mathcal{L}(h)~(h=K,\pi)$ for each hadron hypothesis.
Charged kaons and pions are identified by comparing the likelihoods for the two hypotheses, $\mathcal{L}(K)>\mathcal{L}(\pi)$ and $\mathcal{L}(\pi)>\mathcal{L}(K)$, respectively.

$K^0_S$ candidates are reconstructed with two oppositely charged tracks that satisfy $|V_{z}|<$ 20~cm, assigned as $\pi^+\pi^-$ and without PID imposed.
 Primary vertex and secondary vertex fits are performed on these two charged tracks with $\pi^+\pi^-$ hypothesis to determine the invariant mass and the decay length of $K^0_S$ candidates. 
 The $K^0_S$ candidates are required to have an invariant mass of $\pi^+\pi^-$ ~($M_{\pi^{+}\pi^{-}}$) within [0.487,0.511]GeV/$c^2$, and the decay length to be larger than twice its resolution. 
This requirement is not applied for the $D_s^-\to K_S^0K^-$ decay due to the low combinatorial background. 
To avoid double-counting an event in both the $D_s^-\to K_S^0K^-$ and $D_s^-\to K^-\pi^{-}\pi^{+}$ ST modes, $M_{\pi^{+}\pi^{-}}$ is required to be outside the mass range $[0.487, 0.511]$ GeV$/c^{2}$ for the $D_s^-\to K^-\pi^{-}\pi^{+}$ mode. 

Photon candidates are identified using showers in the EMC. 
The deposited energy of each shower must be more than 25~MeV in the barrel region~($|\!\cos \theta|< 0.80$) and more than 50~MeV in the end-cap region~($0.86 <|\!\cos \theta|< 0.92$). 
To exclude showers that originate from charged tracks,
the opening angle subtended by the EMC shower and the position of the closest charged track at the EMC
must be greater than $10^\circ$ as measured from the IP. 
The difference between the EMC time and the event start time is required to be within [0, 700]\,ns to suppress electronic noise and showers unrelated to the event.

$\pi^0$ candidates are reconstructed through the $\pi^0\to \gamma\gamma$ decay, with at least one photon being detected in the barrel region. 
The invariant mass of the photon pair must be in the range of $[0.115, 0.150]$~GeV/$c^{2}$, approximately three times the mass resolution around the known $\pi^0$ mass~\cite{PDG}. 
A kinematic fit is performed that constrains the $\gamma\gamma$ invariant mass to the known $\pi^{0}$ mass to improve the mass resolution and the $\chi^2$ is required to be less than 30. 

 Five ST modes are taken into account for the reconstruction of $D_s^-$ candidates, with corresponding mass windows on the tagging invariant $D_{s}^{-}$ mass~($M_{\rm tag}$) determined through fits as listed in table~\ref{tab:tag-cut}.
The recoiling mass ($M_{\rm rec}$) against the tagging $D_s^-$ is evaluated, and events with $M_{\rm rec}$ within the mass windows  specified in table~\ref{energe} are retained for further analysis.
The recoiling mass $M_{\rm rec}$ is defined as
\begin{eqnarray}
\begin{aligned}
	\begin{array}{lr}
	M_{\rm rec} = \sqrt{\left(E_{\rm cm}/c^2 - \sqrt{|\vec{p}_{D_{s}^-}/c|^{2}+m_{D_{s}^-}^{2}}\right)^{2} - |\vec{p}_{D_{s}^-}/c | ^{2}} \; , \label{eq:mrec}
		\end{array}
\end{aligned}\end{eqnarray}
where $\vec{p}_{D_{s}^-}$ is the three-momentum of the $D_{s}^{-}$ candidate in the $e^+e^-$ center-of-mass frame, and $m_{D_{s}^-}$ is the known  $D_{s}^{-}$ mass~\cite{PDG}.

\begin{table}[tb!]
\centering
\begin{tabular}{lc}
\hline
Tag mode                                     & Mass window (GeV/$c^{2}$) \\
\hline 
$D_{s}^{-} \to K_{S}^{0}K^{-}$               & [1.948, 1.991]            \\
        $D_{s}^{-} \to K^{+}K^{-}\pi^{-}$            & [1.950, 1.986]            \\
        $D_{s}^{-} \to K^{+}K^{-}\pi^{-}\pi^{0}$     & [1.947, 1.982]            \\
        $D_{s}^{-} \to K_{S}^{0}K^{+}\pi^{-}\pi^{-}$ & [1.953, 1.983]            \\
        $D_{s}^{-} \to K^{-}\pi^{-}\pi^{+}$          & [1.953, 1.983]            \\
\hline
\end{tabular}
\caption{\label{tab:tag-cut} Requirements on $M_{\rm tag}$ for various tag modes.}
\end{table}

\begin{figure*}[htp]
\begin{center}
    \includegraphics[width=6.0cm]{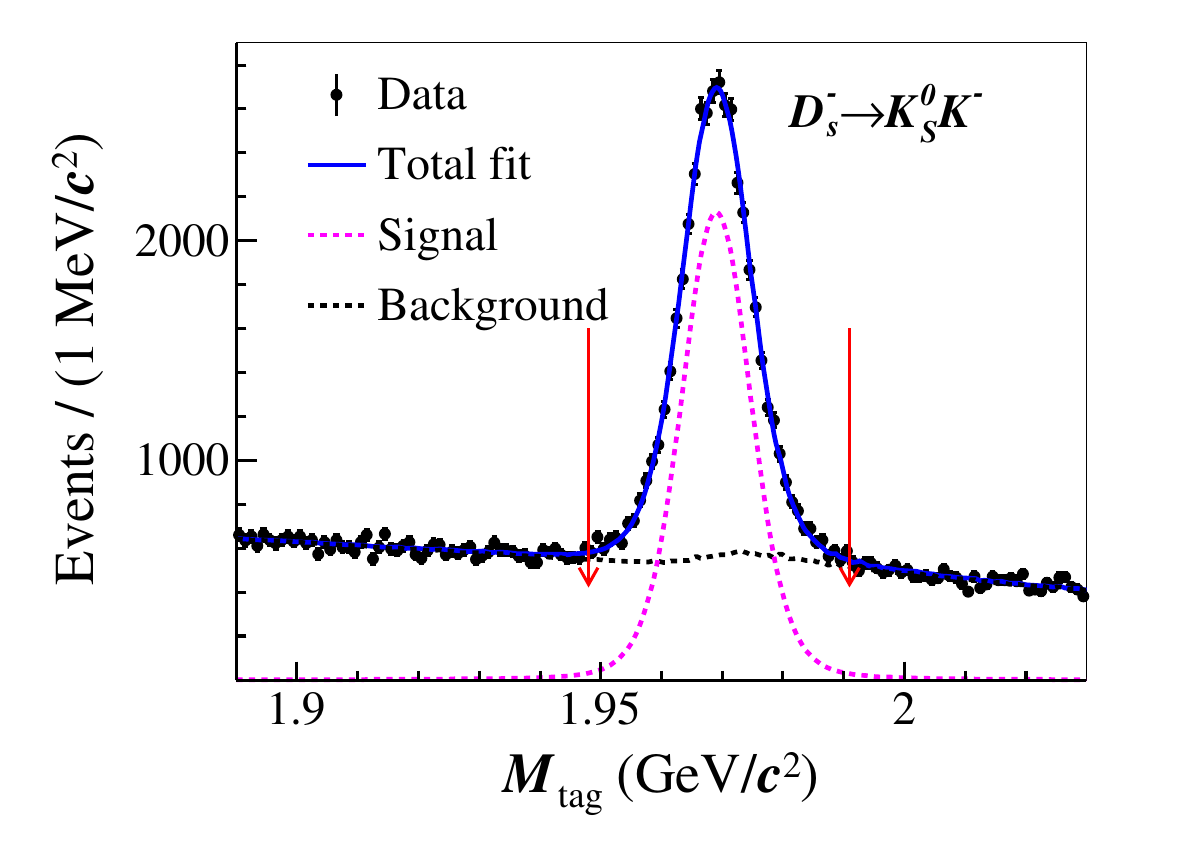}
    \includegraphics[width=6.0cm]{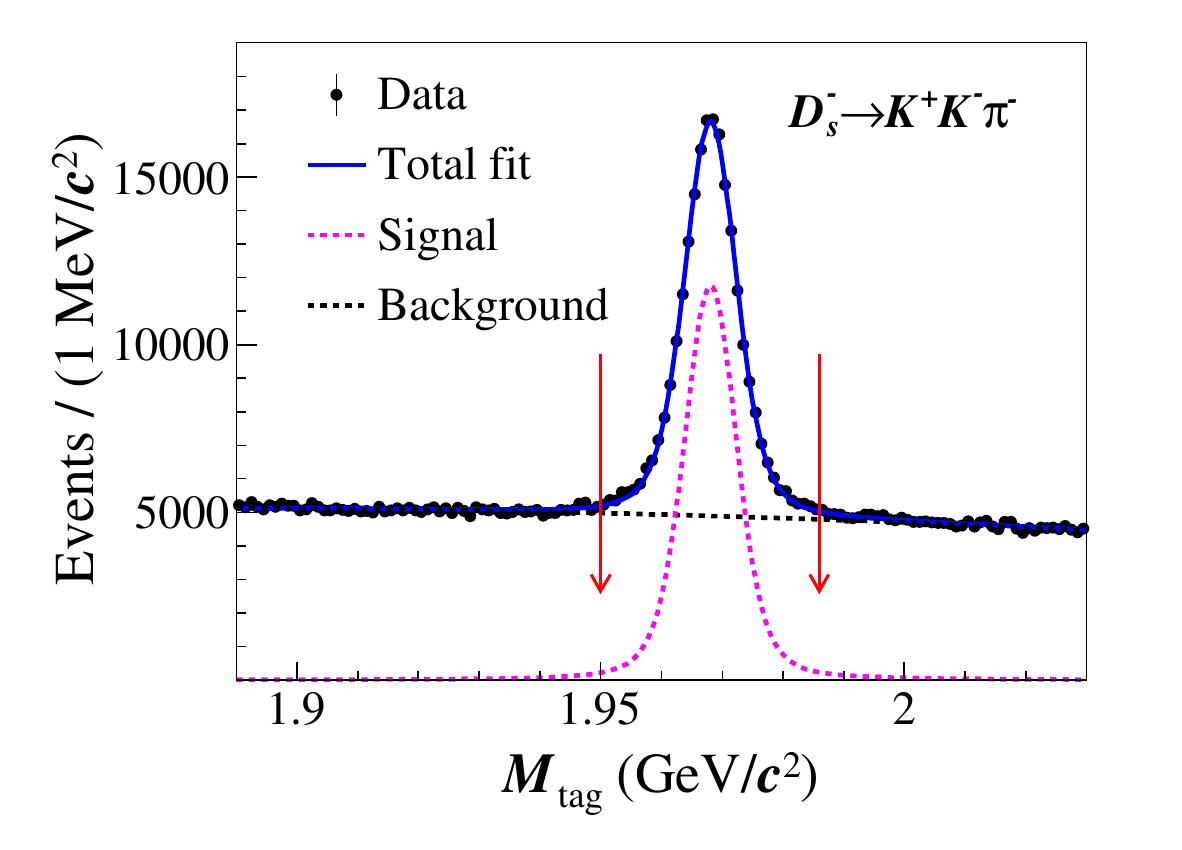}
    \includegraphics[width=6.0cm]{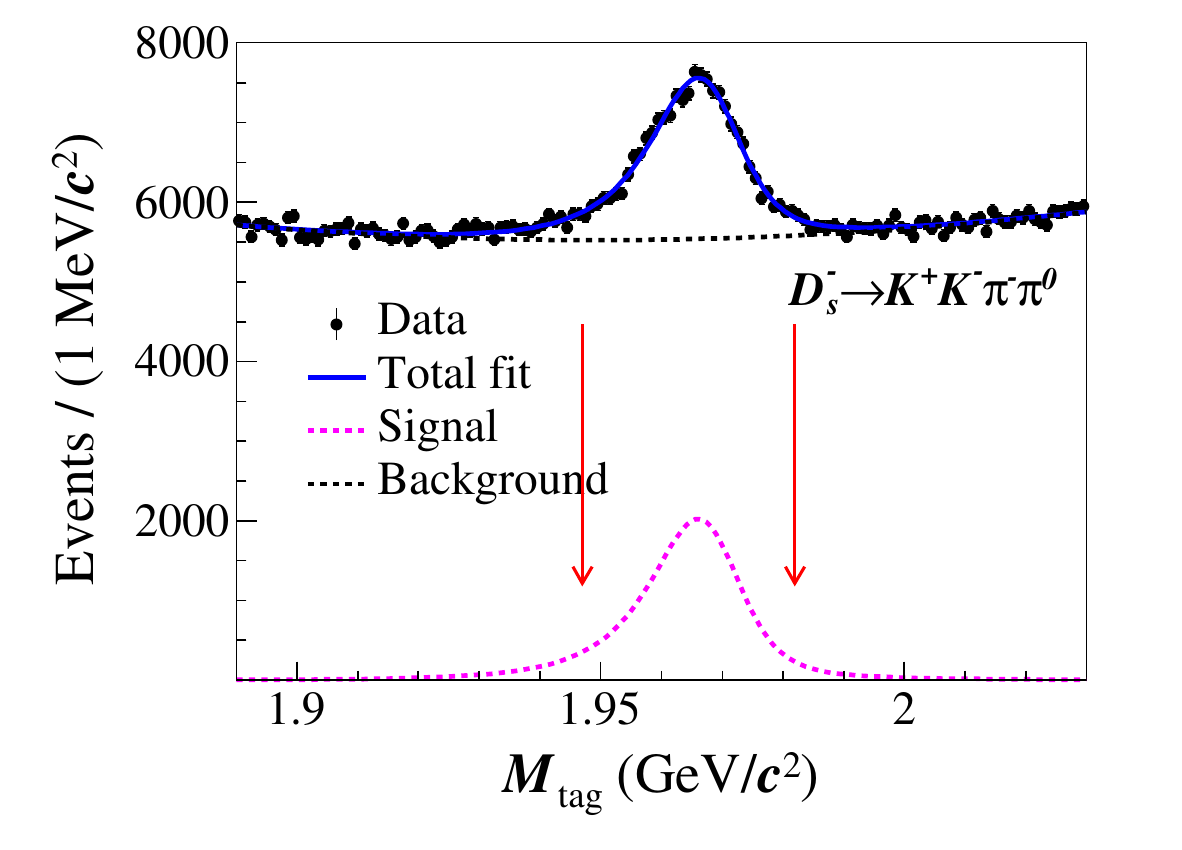}
    \includegraphics[width=6.0cm]{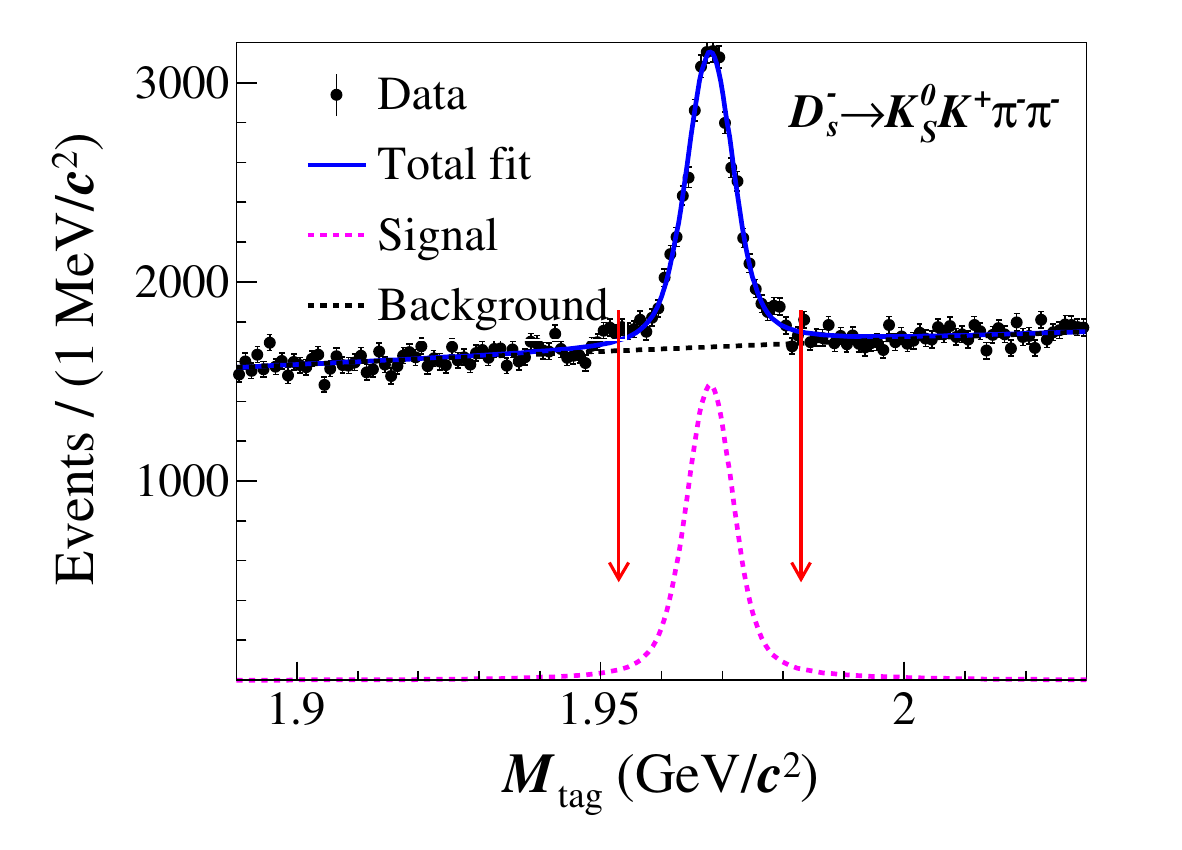}
    \includegraphics[width=6.0cm]{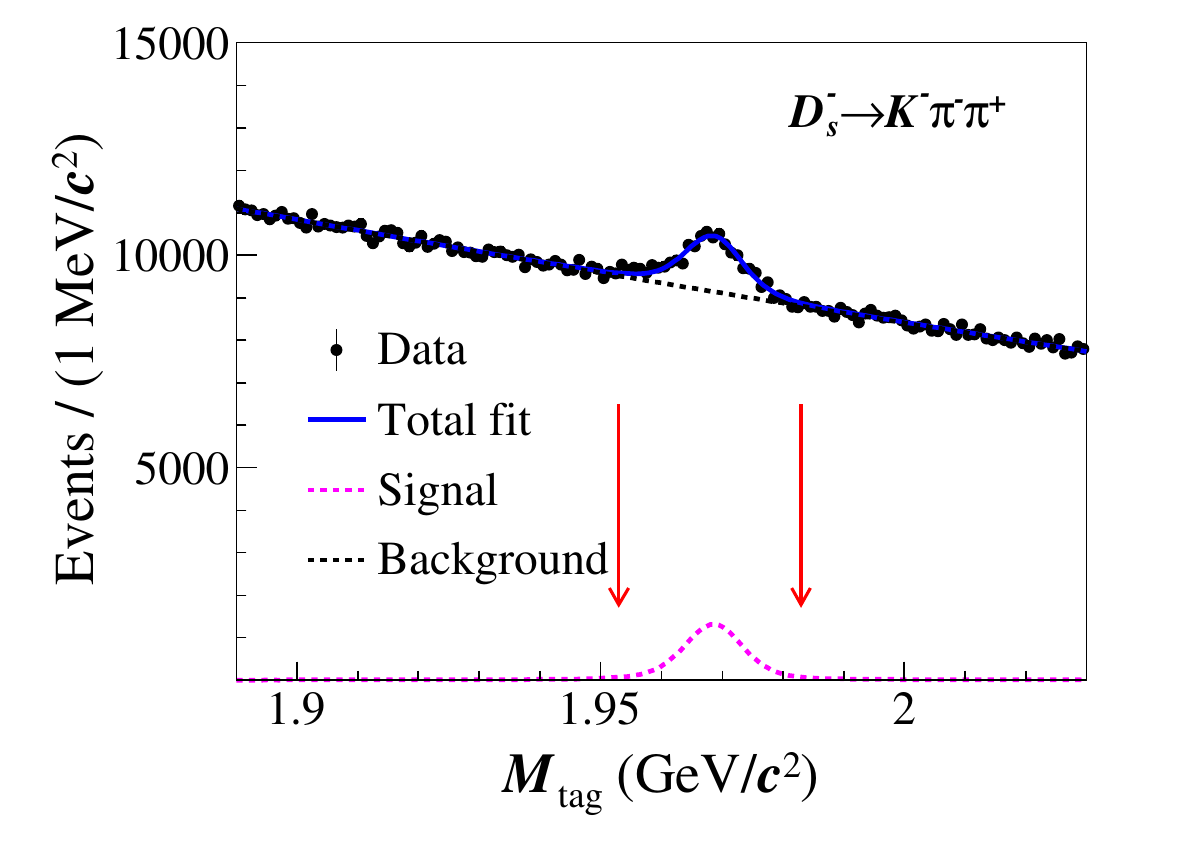}
\caption{Fits to the $M_{\rm tag}$ distributions of the ST candidates from the data sample at $E_{\rm cm} = 4.178$~GeV. 
	The black points with error bars are data, the blue solid lines are the total fit results, the magenta dashed lines show the signal component and the black dashed lines show the background component. 
	The pairs of red arrows denote the signal regions.}
\label{fit:Mass-data-Ds_4180}
\end{center}
\end{figure*}
In order to remove soft pions from $D^{*+}$ decays, the momentum requirement of $p(\pi) > 100$ MeV$/c$ is applied to all pions.
In events with multiple ST candidates, the candidate with $M_{\rm rec}$ closest to the known $D_s^{*+}$ mass~\cite{PDG} is chosen as the best candidate. 
It is observed that approximately 4.1\%(4.3\%) of events in data (inclusive MC) contain multiple candidates in the ST samples.
	The yields for the five ST modes are listed in table~\ref{ST-yield}, and they are obtained through binned maximum likelihoods fits to the corresponding invariant mass distributions of ST candidates~($M_{\rm tag}$). 
As an example, the fits to the accepted ST candidates from the data sample at $E_{\rm cm} = 4.178$~GeV are shown in figure~\ref{fit:Mass-data-Ds_4180}.
In the fits, the signal is modeled by an MC-simulated shape convolved with a Gaussian function that accounts for the data-MC difference. 
The background is described by a second-order Chebyshev polynomial. 
For the tag mode $D_{s}^{-} \to K_{S}^{0} K^-$, the peaking background originating from $D^{-} \to K_{S}^{0} \pi^-$ is considered.
	The shape of this background is taken from the inclusive MC samples and included in the fit with a free yield parameter. 
	The same ST selection criteria and fitting approach applied to data are used to analyze the inclusive MC samples. The number of observed ST events is extracted from
fitting the $M_{\rm tag}$ distributions. 
The ST efficiency is calculated as the ratio of the observed ST events and the generated ST events in the inclusive MC samples, as shown in  table~\ref{ST-eff}. 
When obtaining the ST efficiencies within a grouped dataset~(I and III), the yields at different energy points are averaged based on their luminosity and cross sections.
\begin{table*}[htp]
	\centering
  \begin{tabular}{l r@{$\pm$}l r@{$\pm$}l r@{$\pm$}l r@{$\pm$}l }
  \hline
    Tag mode   &  \multicolumn{2}{c}{$N^{\text{ST}}_{\rm I}$}  
    &\multicolumn{2}{c}{$N^{\text{ST}}_{\rm II}$}   
		&\multicolumn{2}{c}{$N^{\text{ST}}_{\rm III}$}     
		&\multicolumn{2}{c}{$N^{\text{ST}}_{\rm IV}$}   \\
  \hline
    $D^-_s\to K^0_SK^-$          &6728  &144  &31949  &314  &19960 &270  &6837  &163\\
    $D^-_s\to K^+K^-\pi^-$       &27670 &280  &137138 &614  &86918 &525  &29544 &335  \\
    $D^-_s\to K^+K^-\pi^-\pi^0$  &7457  &397  &39340  &798  &24694 &688  &8084  &481 \\
    $D^-_s\to K_S^0K^+\pi^-\pi^-$&2983  &129  &15705  &288  &9783  &247  &3380  &174\\
    $D^-_s\to K^-\pi^-\pi^+$     &3804  &345  &17439  &565  &10841 &470  &5144  &447\\
    \hline
  \end{tabular}\caption{The ST yields~($N^{\text{ST}}_\alpha$) for the data samples collected at $E_{\rm cm} =$ 4.128 and 4.157~GeV (I), 4.178~GeV (II), from 4.189 to 4.219~GeV (III), 4.226~GeV (IV). The uncertainties are statistical.}
  \label{ST-yield}
\end{table*}

\begin{table*}[htp]
	\centering
  \begin{tabular}{l r@{$\pm$}l r@{$\pm$}l r@{$\pm$}l r@{$\pm$}l}
  \hline
    Tag mode   &\multicolumn{2}{c}{$\epsilon^{\rm ST}_{\rm I} (\%)$} 
    &\multicolumn{2}{c}{$\epsilon^{\rm ST}_{\rm II} (\%)$}
		&\multicolumn{2}{c}{$\epsilon^{\rm ST}_{\rm III} (\%)$}  
		&\multicolumn{2}{c}{$\epsilon^{\rm ST}_{\rm IV} (\%)$}\\
  \hline
    $D^-_s\to K^0_SK^-$          &47.64 &0.16 &47.39 &0.07 &47.23 &0.09  &47.95 &0.16\\
    $D^-_s\to K^+K^-\pi^-$       &40.37 &0.07 &39.47 &0.03 &39.33 &0.04  &39.78 &0.07\\
    $D^-_s\to K^+K^-\pi^-\pi^0$  &10.59 &0.08 &10.68 &0.03 &10.74 &0.05  &10.89 &0.09\\
    $D^-_s\to K_S^0K^+\pi^-\pi^-$&21.30 &0.14 &21.85 &0.06 &21.66 &0.08  &22.27 &0.16\\
    $D^-_s\to K^-\pi^-\pi^+$     &48.37 &0.59 &47.93 &0.25 &47.63 &0.34  &47.67 &0.67\\
    \hline
  \end{tabular}
  \caption{The ST efficiencies~($\epsilon^{\rm ST}_{\rm \alpha}$) for the data samples collected at $E_{\rm cm} =$ 4.128 and 4.157~GeV (I), 4.178~GeV (II), from 4.189 to 4.219~GeV (III), 4.226~GeV (IV). The uncertainties are statistical.}
  \label{ST-eff}
\end{table*}

\begin{figure}[tp!]
          \centering
          \includegraphics[width=7.5cm]{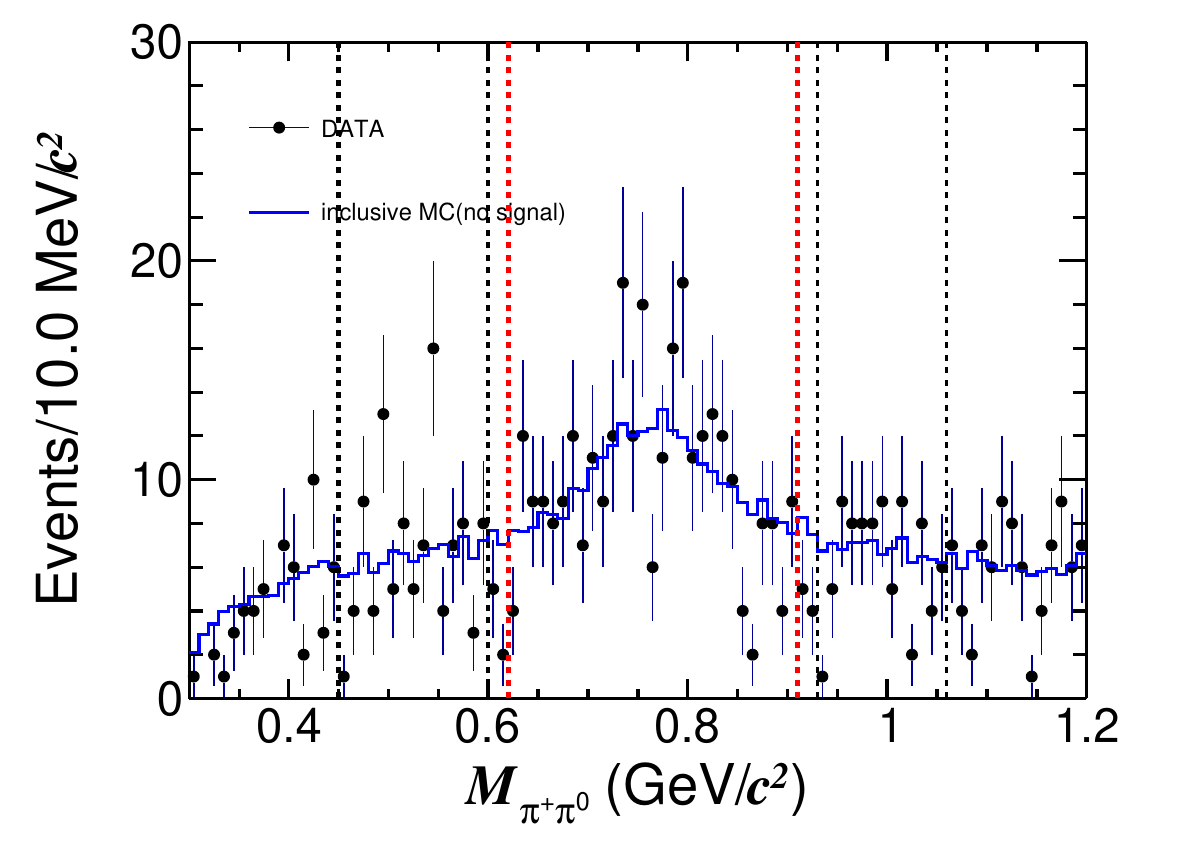}
    \caption{Invariant mass distribution for selected $\rho(770)^+\to\pi^+\pi^0$ candidates at $\sqrt{s}=4.128-4.226$ GeV.
    The points with error bars are data.
    The red vertical dashed lines indicate the $\rho(770)^+$ signal region, and the pairs of black dashed lines (left and right of the signal peak) indicate the $\rho(770)^+$ sideband regions.}
    \label{omega}
\end{figure}

In events with a ST $D_s^-$ candidate, we search for the $D^+_s\to \gamma\rho(770)^+$ signal process recoiling against the ST side, with the $\rho(770)^+$ reconstructed with the $\pi^+\pi^0$ final states.
The invariant mass distribution of $\pi^+\pi^0$ for data and inclusive MC sample is shown in Fig.~\ref{omega}.
There is no obvious background from $D_s^+ \to \gamma \pi^+\pi^0$ in the $\rho(770)^+$ sideband regions ([0.45,0.60] and [0.93,1.06] GeV/$c^2$).
Therefore we require the $\pi^+\pi^0$ invariant mass to be in the $\rho(770)^+$ signal range [0.62, 0.91] GeV/$c^2$ for further analysis.
In the case of multiple candidates, the DT candidate with the average mass, $(M_{\rm sig}+M_{\rm tag})/2$, closest to the known $D_{s}^{\pm}$ mass is retained, where $M_{\rm sig}$ represents the invariant mass of the accepted signal $D^+_s$. 
For all ST modes, the average ratio of correct selection for multiple signal candidates is found to be $94\%$.

To suppress background events from $D_s^+\to\pi^+\eta$, $\eta\to\gamma\gamma$, a requirement of $M_{\gamma\gamma_h}>0.68$ GeV/$c^2$ is imposed to ensure the invariant mass is above the $\eta$ invariant mass, where $\gamma_h$ is the photon with higher energy from the $\pi^0$ of the $\rho^+$ decay in the signal reaction and $\gamma$ is the radiative photon from the signal process $D_s^+\to \gamma \rho^+$.
These selection criteria are optimized using the figure of merit~(FOM) approach, defined as $\frac{\epsilon}{1.5+\sqrt{B}}$~\cite{Punzi:2020fsv}. Here, $\epsilon$ and $B$ denote the signal efficiency estimated by the signal MC samples and the background yields estimated by the inclusive MC samples, respectively.
If the invariant mass of $M_{\gamma\gamma_{\rm extra}}$ falls within the $\pi^0$ mass window [0.115, 0.150] GeV/$c^2$ or the $\eta$ mass window [0.50, 0.57] GeV/$c^2$, such candidates are vetoed, with $\gamma_{\rm extra}$ denoting an extra photon not utilized in either the ST or DT sides.. 

\section{Branching fraction measurement}
\label{BFSelection} 
\begin{figure}[tp!]
\begin{center}
      \includegraphics[width=4.95cm]{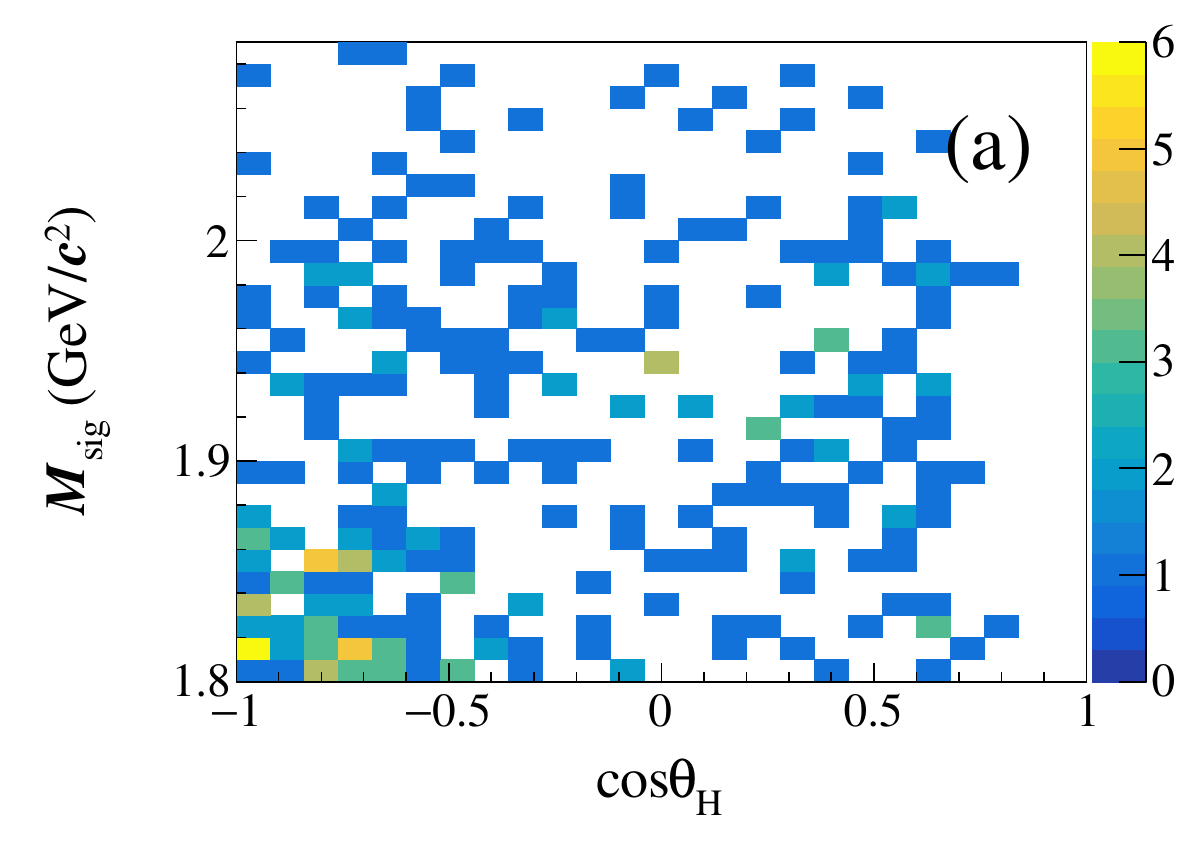}
      \includegraphics[width=4.95cm]{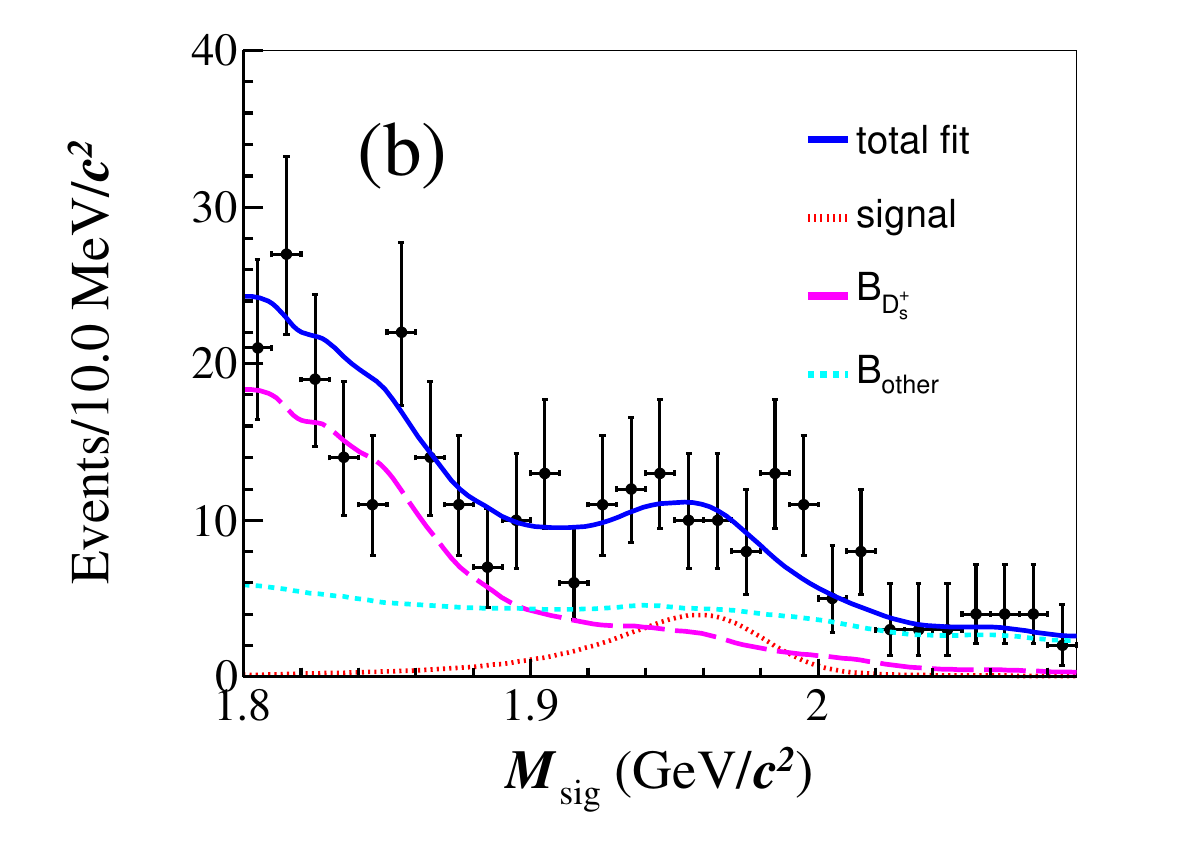}
      \includegraphics[width=4.95cm]{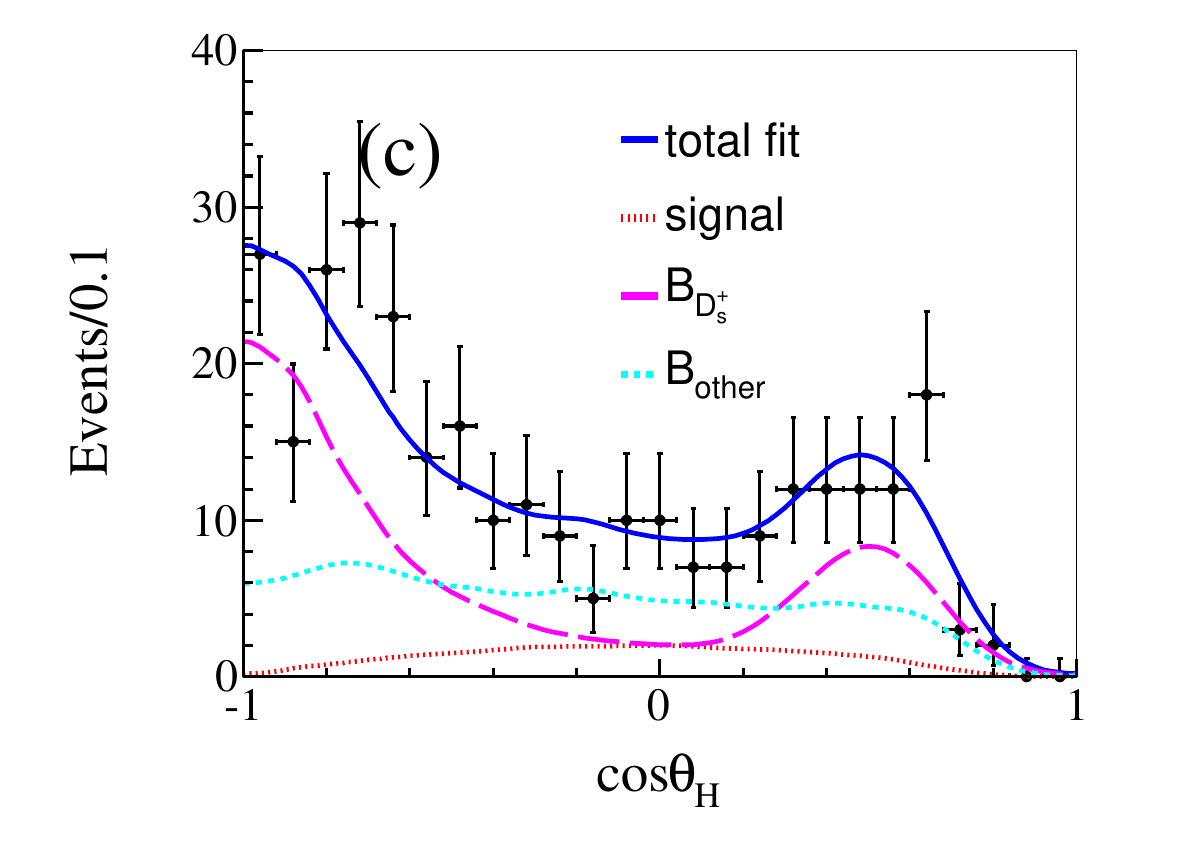}
   \caption{Distributions of $M_{\rm sig}$ versus cos$\theta_{\rm H}$~(a), $M_{\rm sig}$~(b) and cos$\theta_{\rm H}$~(c) for the DT candidate events of $D^+_s\to\gamma\rho(770)^+$ at $E_{\rm cm} = 4128-4226$ MeV.
	The black points with error bars are data, the blue solid curves are the total fit result, the red dotted lines are the signal contribution, the magenta long dashed lines illustrate $\rm B_{D_s^{+}}$, and the cyan dashed lines illustrate $\rm B_{\rm other}$.}
\label{2d}
\end{center}
\end{figure}

After imposing all above selection criteria, the distributions of the invariant mass ($M_{\rm {sig}}$) of signal $D_s^+$ candidates versus cos$\theta_{\rm H}$, $M_{\rm {sig}}$ and cos$\theta_{\rm H}$ are shown in the figures.~\ref{2d}~(a), (b), and (c), respectively.
The signal yield is extracted from a two-dimensional~(2D) unbinned maximum likelihood fit on the $M_{\rm sig}$ versus cos$\theta_{\rm H}$ distribution for $D_s^+\to\gamma\rho(770)^+$. 
The signal shape is described by an MC-simulated 2D probability density function (PDF) convolved with a 2D Gaussian function, with the parameters obtained from the control sample $D^{+}_{s}\to \pi^+\pi^0\eta$~\cite{BESIII:2019jjr}.
The yield of background events originating from $D^{+}_{s}D_s^{-}$ pairs ($\rm B_{D_s^{+}}$) is constraint to a Gaussian in the fit, with the mean and resolution of the Gaussian fixed at $166\pm9$ determined based on MC simulation.
The shape of the background components of non-$D_s^+D_s^-$ reactions ($\rm B_{\rm other}$) is derived from the MC simulated events, and its yield is free in the fit.

\begin{table}[thp]
  \centering
  \begin{tabular}{l r@{$\pm$}l r@{$\pm$}l r@{$\pm$}l r@{$\pm$}l}
  \hline
    Tag mode   & \multicolumn{2}{c}{$\epsilon^{\rm DT}_{\rm I} (\%)$} &\multicolumn{2}{c}{$\epsilon^{\rm DT}_{\rm II} (\%)$}    &\multicolumn{2}{c}{$\epsilon^{\rm DT}_{\rm III} (\%)$} &\multicolumn{2}{c}{$\epsilon^{\rm DT}_{\rm IV} (\%)$}\\
  \hline
    $D^-_s\to K^0_SK^-$           &14.39 &0.28 &14.68 &0.13 &14.29 &0.16 &14.66 &0.31\\
    $D^-_s\to K^+K^-\pi^-$        &11.54 &0.11 &11.79 &0.05 &11.52 &0.06 &11.72 &0.12\\
    $D^-_s\to K^+K^-\pi^-\pi^0$   &3.38 &0.06  &3.41 &0.02  &3.42 &0.03  &3.51 &0.06\\
    $D^-_s\to K^+_s K^+\pi^-\pi^-$&5.77 &0.17  &6.21 &0.08  &6.24 &0.09  &6.61 &0.20\\
    $D^-_s\to K^-\pi^-\pi^+$      &14.24 &0.37 &13.38 &0.15 &13.16 &0.18 &13.82 &0.37\\
    \hline
  \end{tabular}
	\caption{The DT efficiencies~($\epsilon^{\rm DT}_{\alpha}$) for the data samples taken at $E_{\rm cm} = 4.128$ and $4.157$ GeV (I), 4.178 GeV (II), from 4.189 to 4.219 GeV (III), and 4.226 GeV (IV). }
  \label{abs:dteff}
\end{table}

From the 2D fit, the number of signal events is determined to be $33\pm14$, 
with a statistical significance of 2.5$\sigma$. The statistical significance of the $D^+_s\to\gamma\rho(770)^+$ decay is evaluated using $\sqrt{-2{\rm ln}(\mathcal {L}_0/\mathcal {L}_{\rm max})}$, where $\mathcal {L}_{\rm max}$ is the maximum likelihood of the nominal fit and $\mathcal {L}_0$ is the likelihood of the fit without the signal component.
The corresponding DT efficiencies for each tag mode are determined and listed in table~\ref{abs:dteff}. 
The BF of $D^+_s\to\gamma\rho(770)^+$ is determined to be $(2.2\pm0.9)\times10^{-4}$, where the uncertainty is statistical.

\section{Systematic uncertainty}
The systematic uncertainties in the BF measurement are discussed below.
The additive systematic uncertainties only affect the fitted signal yield, while multiplicative systematic uncertainties affect the detection efficiency. 
Most systematic uncertainties related to the determination of the efficiency for reconstructing the ST side cancel out in the BF measurement due to the DT technique. 

At first, the multiplicative systematic uncertainties are discussed below.
\begin{itemize}
\item ST yield. 
The total ST yield of all five ST modes is $495,398\pm1,900$,
		resulting in the statistical uncertainty of $\sqrt{1,900^2-495,398}/495,398 = 0.4\%$.
Here, we only consider the statistical fluctuation related to the background of the ST side,
as it is not correlated with the DT sample directly. 
 Hence, a systematic uncertainty of $0.4\%$ is assigned.

\item Tracking and PID. The $\pi^+$ tracking and PID efficiencies are studied with $e^+e^-\to K^+K^-\pi^+\pi^-$ events. 
The systematic uncertainties are assigned as the data-MC differences of the $\pi^+$ PID and tracking efficiencies, which are both $1.0\%$~\cite{BESIII:2020ctr}.

\item $\gamma$ reconstruction. The uncertainty associated with the $\gamma$ reconstruction efficiency
is studied with the control sample of $J/\psi \to \pi^0\pi^+\pi^-$~\cite{BESIII:2011ysp}. The uncertainty due to the $\gamma$ reconstruction is assigned as $1.0\%$.

\item $\pi^0$ reconstruction. The systematic uncertainty associated with the $\pi^0$ reconstruction efficiency is investigated by using a control sample of the process $e^+e^-\to K^+K^-\pi^+\pi^-\pi^0$~\cite{BESIII:2023mie}. 
The uncertainty due to the $\pi^0$ reconstruction is assigned as $2.0\%$.

\item MC statistics. The uncertainty due to the limited signal MC sample size is obtained by $\sqrt{\sum_i{(\frac{f_i\delta_{\epsilon_i}}{\epsilon_i})^2}}$, where $f_i$ is the proportion of each ST yield to the total ST yield, and $\epsilon_i$ and $\delta_{\epsilon_i}$ are the signal efficiency and the corresponding uncertainty of ST mode $i$, respectively. 
This uncertainty is found to be $0.4\%$.

\item $M_{\gamma\gamma_{\rm h}}$ requirement and $M_{\gamma\gamma\rm extra}$ requirement. The systematic uncertainties due to the $M_{\gamma\gamma_{\rm h}}$ requirement and $M_{\gamma\gamma\rm extra}$ requirement are studied using a control sample of $D_s^+\to\pi^+\pi^0 \eta$~\cite{BESIII:2019jjr}. 
The differences in the efficiencies of the $M_{\gamma\gamma_{\rm h}}$ requirement and $M_{\gamma\gamma\rm extra}$ requirement between data and MC simulation, $1.5\%$ and $0.9\%$, are taken as the corresponding systematic uncertainties.

\item $\rho^+$ mass window. The systematic uncertainties due to the $\rho^+$ mass window is studied using a control sample of $D^0\to\pi^-\pi^0 e^+ \nu_e$~\cite{BESIII:2018qmf}. 
The differences in the efficiencies of the $\rho^+$ mass window between data and MC simulation, $1.0\%$, are taken as the corresponding systematic uncertainties.


\end{itemize}

The additive systematic uncertainties are discussed below. 
\begin{itemize}
\item Signal shape resolution. To estimate the systematic uncertainty related with the signal shape resolution, we examine the BF by varying the mean and resolution of the convolving Gaussian function within their corresponding uncertainties.
We take the change of the BF, 6.4$\%$, as the systematic uncertainty. 
Among the results of these fits, the one corresponding to the largest upper limit on the branching fraction is chosen.

\item Background shape. 
We vary the yield of the $\rm B_{D_s^{+}}$ within corresponding uncertainty and assign the change of the BF of 1.2$\%$ as the systematic uncertainty.
For $\rm B_{\rm other}$, we alter the MC-simulated shapes by varying the relative fractions of the two major background components from $q\bar q$ and non-$D_s^{*+}D_s^-$ open charm by $\pm 30\%$~\cite{BESIII:2021xox}, which is the statistical uncertainty of the individual cross section in data. 
The largest change of the BF, 7.2$\%$, is taken as the systematic uncertainty. 
Among the variations, the fit resulting in the largest upper limit on the branching fraction is chosen.
\end{itemize}
The systematic uncertainties are summarised in table~\ref{abs:bfsys}.
		Adding them in quadrature gives a total systematic uncertainty of the BF measurement of $10.5\%$.


\begin{table}[htbp]
  \centering
  \begin{tabular}{l|l|c}
  \hline
    \hline
    Category &Source   &Uncertainty (\%) \\
  \hline
    \multirow{10}{*}{Multiplicative}&ST yield      &0.4\\
    &Tracking                                &1.0\\
    &PID                                     &1.0\\
    &$\pi^0$ and $\gamma$ reconstructions     &3.0\\
    &MC statistics                           &0.4\\
    &$M_{\gamma\gamma_{\rm h}}$ requirement     &1.5\\
    &$M_{\gamma\gamma_{\rm extra}}$ requirement &0.9\\
    &$\rho(770)^+$ mass window                     &1.0\\
    \cline{2-3}
    &Total                                   &3.9\\
    \hline
    \multirow{4}{*}{Additive} &Signal shape resolution &6.4\\
    &Background $\rm B_{D_s^{+}}$                           &1.2\\
    &Background $\rm B_{\rm other}$                          &7.2\\
    \cline{2-3}
    &Total                                   &9.7\\
    \hline
    \hline
  \end{tabular}
	\caption{Systematic uncertainties in the BF measurement of $D^+_s\to\gamma\rho(770)^+$.}
  \label{abs:bfsys}
\end{table}

\begin{figure}[htbp]
          \centering
          \includegraphics[width=10.0cm]{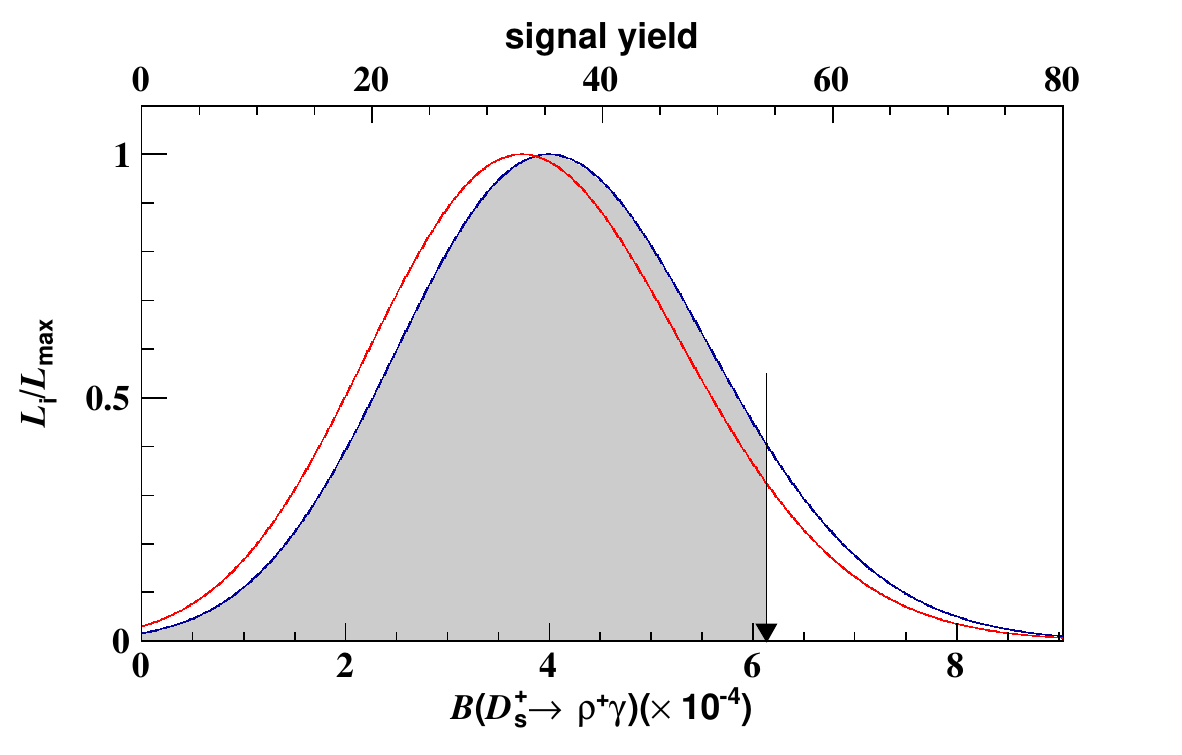}
		\caption{
		Normalized likelihood versus the signal yield and BF of $D^+_s\to\gamma\rho(770)^+$. 
		The red solid line is the likelihood curve for the nominal fit model, while the blue solid line represents the likelihood curve that gives the upper limit after incorporating the systematic uncertainty. 
		The black arrow indicates the result corresponding to the $90\%$ confidence level.
		}
    \label{fig:bf2}
\end{figure}

The absolute BF of $D^+_s\to \gamma\rho(770)^+$ is measured to be $(2.2\pm0.9\pm0.2)\times10^{-4}$, where the first uncertainty is statistical and the second is systematic.
Because of the limited statistics, an upper limit on the BF of $D^+_s\to\gamma\rho(770)^+$ is also determined following Ref.~\cite{BESIII:2022jcm}, after incorporating the systematic uncertainty via a likelihood scan method. 
To take into account the additive systematic uncertainties, the maximum-likelihood fits are repeated from the $\rm B_{\rm other}$ shapes as mentioned in the previous section and the one resulting in the most conservative upper limit is chosen. 
Finally, the multiplicative systematic uncertainty $\sigma_{\epsilon}$ is incorporated in the
calculation of the upper limit following Refs.~\cite{Stenson:2006gwf, Liu:2015uha}.
\begin{equation}
	L({\cal B})\propto\int_0^1L({\cal B} \cdot \frac{\epsilon}{\epsilon_0})e^{{-\frac{(\epsilon-\epsilon_0)^2}{2(\sigma_\epsilon\epsilon_0)^2}}}d\epsilon, 
\end{equation}
where $L({\cal B})$ is the likelihood distribution as a function of assumed BFs; $\epsilon$ is the expected efficiency and $\epsilon_0$ is the averaged MC-estimated efficiency. 
The likelihood distributions with and without incorporating the systematic uncertainties are shown in figure~\ref{fig:bf2}. 
The upper limit of the BF of $D^+_s\to\gamma\rho(770)^+$ is obtained to be $6.1\times10^{-4}$ at the $90\%$ confidence level.

\section{Summary}
\begin{figure}[tp!]
\begin{center}
        \includegraphics[width=10.0cm]{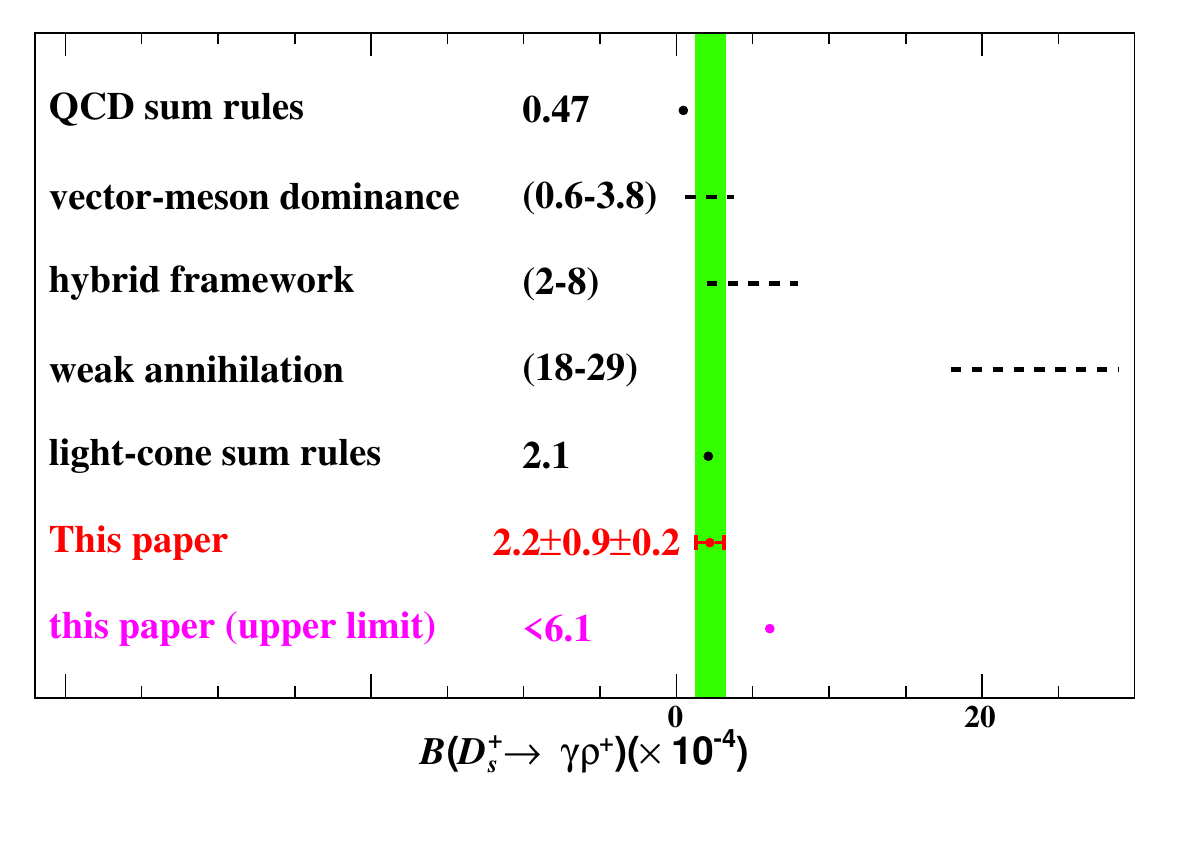}
   \caption{The comparison of ${\cal B}(D^+_s\to \gamma\rho(770)^+)$ obtained in this
work with theoretical calculations. 
The green band denotes the experimental result of this paper.}
\label{fig:bf}
\end{center}
\end{figure}

In summary, using 7.33~fb$^{-1}$ of $e^+e^-$ collision data collected with the BESIII detector between $E_{\rm cm} = 4.128$ and $4.226$ GeV, we search for the rare radiative decay $D^+_s\to\gamma\rho(770)^+$ for the first time. 
A hint with a statistical significance of $2.5\sigma$ is obtained. 
The upper limit on the BF for $D_s^+\to\gamma\rho(770)^+$ is estimated to be $\mathcal{B}(D^+_s\to \gamma\rho(770)^+) < 6.1\times$10$^{-4}$ at the $90\%$ confidence level, as shown in figure~\ref{fig:bf2}. 
The absolute BF of this decay is measured to be ${\cal B}(D^+_s\to \gamma\rho(770)^+) = (2.2\pm0.9\pm0.2)\times10^{-4}$, where the first uncertainty is statistical and the second is systematic. 
Figure~\ref{fig:bf} shows the comparison of our measurement with different theoretical predicted BFs.
The obtained result in this analysis agrees with theoretical calculations based on light-cone sum rules~\cite{Lyon:2012fk}, hybrid framework~\cite{Fajfer:1998dv,Fajfer:1997bh} and vector-meson dominance~\cite{Burdman:1995te}. 
However, it is 3.8$\sigma$ smaller than the theoretical calculation of weak annihilation~\cite{deBoer:2017que}.
This study is beneficial for further understanding of the non-perturbative QCD in the $D^+_s$ sector.
 
\acknowledgments
The BESIII Collaboration thanks the staff of BEPCII and the IHEP computing center for their strong support. This work is supported in part by National Key R\&D Program of China under Contracts Nos. 2020YFA0406400, 2020YFA0406300, 2023YFA1606000; National Natural Science Foundation of China (NSFC) under Contracts Nos. 11635010, 11735014, 11935015, 11935016, 11935018, 12025502, 12035009, 12035013, 12061131003, 12192260, 12192261, 12192262, 12192263, 12192264, 12192265, 12221005, 12225509, 12235017, 12361141819; the Chinese Academy of Sciences (CAS) Large-Scale Scientific Facility Program; the CAS Center for Excellence in Particle Physics (CCEPP); Joint Large-Scale Scientific Facility Funds of the NSFC and CAS under Contract No. U1832207, U1932108, U2032104; 100 Talents Program of CAS; The Institute of Nuclear and Particle Physics (INPAC) and Shanghai Key Laboratory for Particle Physics and Cosmology; German Research Foundation DFG under Contracts Nos. 455635585, FOR5327, GRK 2149; Istituto Nazionale di Fisica Nucleare, Italy; Knut and Alice Wallenberg Foundation under Contracts Nos. 2021.0174, 2021.0299; Ministry of Development of Turkey under Contract No. DPT2006K-120470; National Research Foundation of Korea under Contract No. NRF-2022R1A2C1092335; National Science and Technology fund of Mongolia; National Science Research and Innovation Fund (NSRF) via the Program Management Unit for Human Resources \& Institutional Development, Research and Innovation of Thailand under Contracts Nos. B16F640076, B50G670107; Polish National Science Centre under Contract No. 2019/35/O/ST2/02907; The Swedish Research Council; U. S. Department of Energy under Contract No. DE-FG02-05ER41374

\bibliographystyle{JHEP}
\bibliography{references}

\clearpage
\appendix
M.~Ablikim$^{1}$, M.~N.~Achasov$^{4,c}$, P.~Adlarson$^{76}$, O.~Afedulidis$^{3}$, X.~C.~Ai$^{81}$, R.~Aliberti$^{35}$, A.~Amoroso$^{75A,75C}$, Q.~An$^{72,58,a}$, Y.~Bai$^{57}$, O.~Bakina$^{36}$
% ...
\\
\vspace{0.2cm}
(BESIII Collaboration)\\
\vspace{0.2cm} {\it
$^{1}$ Institute of High Energy Physics, Beijing 100049, People's Republic of China\\
% ...
$^{3}$ Bochum  Ruhr-University, D-44780 Bochum, Germany\\
$^{4}$ Budker Institute of Nuclear Physics SB RAS (BINP), Novosibirsk 630090, Russia\\
% ...
$^{35}$ Johannes Gutenberg University of Mainz, Johann-Joachim-Becher-Weg 45, D-55099 Mainz, Germany\\
$^{36}$ Joint Institute for Nuclear Research, 141980 Dubna, Moscow region, Russia\\
% ...
$^{57}$ Southeast University, Nanjing 211100, People's Republic of China\\
$^{58}$ State Key Laboratory of Particle Detection and Electronics, Beijing 100049, Hefei 230026, People's Republic of China\\
% ...
$^{72}$ University of Science and Technology of China, Hefei 230026, People's Republic of China\\
% ...
$^{75}$ University of Turin and INFN, (A)University of Turin, I-10125, Turin, Italy; (B)University of Eastern Piedmont, I-15121, Alessandria, Italy; (C)INFN, I-10125, Turin, Italy\\
$^{76}$ Uppsala University, Box 516, SE-75120 Uppsala, Sweden\\
% ...
$^{81}$ Zhengzhou University, Zhengzhou 450001, People's Republic of China\\

\vspace{0.2cm}
$^{a}$ Deceased\\
% ...
$^{c}$ Also at the Novosibirsk State University, Novosibirsk, 630090, Russia\\
% ...

}

\end{document}